\documentclass[prl,twocolumn,graphicx,preprintnumbers, showpacs]{revtex4-1}
\usepackage{epsfig}
\usepackage{lineno} 
\usepackage{natbib}
\usepackage{dcolumn}
\usepackage{subfigure}
\usepackage{hyperref} 

\newcommand{\pt}{\mbox{\ensuremath{p_{\rm T}}}}
\newcommand{\met}{\mbox{\ensuremath{E_{\rm T}^{\rm miss}}}}

\begin{document}

\preprint{CERN-PH-EP-2012-007}
\preprint{Submitted to Physical Review Letters}
\title{\vspace*{0.3cm}Search for Pair Production of a Heavy Up-Type Quark Decaying to a $W$ Boson \\ and a $b$ Quark in the Lepton+Jets Channel with the ATLAS Detector}

\author{ATLAS Collaboration}

\affiliation{}

\date{\today}

\begin{abstract}
A search is presented for production of a heavy up-type quark ($t^\prime$) together with its antiparticle, assuming subsequent decay 
to a $W$ boson and a $b$ quark,  $t^\prime \bar{t^\prime} \to W^+b W^- \bar{b}$. The search is
based on 1.04~fb$^{-1}$ of proton-proton collisions at $\sqrt{s}=7$~TeV collected by the ATLAS detector at the CERN Large Hadron Collider.
Data are analyzed in the lepton+jets final state, characterized by a high transverse momentum isolated electron or muon, high missing
transverse momentum and at least three jets.
No significant excess of events above the background expectation is observed. 
A  95\% C.L. lower limit of 404~GeV is set for the mass of the $t^\prime$ quark.
\end{abstract}


\maketitle 

The discovery of the top quark~\cite{topdisc_cdf,*topdisc_d0} completed the third generation of fundamental fermions in the
quark sector of the Standard Model (SM) of particle physics.
It is natural to ask whether heavier quarks may exist.
These quarks are often present in new physics models aimed at solving the limitations of the SM. For example, models with a fourth generation of heavy chiral fermions could provide new sources of CP 
violation to explain the matter-antimatter asymmetry in the Universe, and 
allow for a heavier Higgs boson while remaining consistent with precision electroweak data~\cite{*[{See e.g. }][{ and references therein.}] holdom}. 
The latter is accomplished by keeping a small mass splitting between the heavy up-type quark ($t^\prime$)
and the heavy down-type quark ($b^\prime$). Assuming that $m_{t^\prime}-m_{b^\prime}<m_W$, where $m_W$ is the $W$ boson mass, results
in the $t^\prime$ quark predominantly decaying to a $W$ boson and a down-type quark $q$ ($q$=$d$, $s$, $b$).
Another possibility is the addition of isospin singlets or doublets of vector-like quarks, which appear in many extensions of the SM such as 
Little Higgs or extra-dimensional models~\cite{*[{See e.g. }][{ and references therein.}] jaas}.
In both scenarios the $t^\prime$ quark can decay into  $Wb$  with
a large branching ratio, provided there is a significant mixing with the third generation of quarks, 
consistent with the existing mass and mixing patterns of the known quarks.

The high center-of-mass energy and integrated luminosity in $pp$ collisions available at the Large Hadron Collider (LHC) offers
a unique opportunity to probe these scenarios.
At the LHC,  these new heavy quarks would be predominantly produced in pairs via the strong 
interaction for masses below $\sim 1$~TeV, while for larger masses electroweak production of single heavy quarks could become
the primary production mechanism, depending on the strength of their interactions with the SM quarks and weak gauge bosons~\cite{jaas}.

A search is presented in this Letter for $t^\prime \bar{t^\prime}$ production using $pp$ collision data at $\sqrt{s}=7$~TeV 
collected with the ATLAS detector. 
It is assumed that the $t^\prime$ quark decays exclusively into $Wb$. 
The lepton+jets final state signature is considered,  characterized by a high transverse momentum ($\pt$) isolated electron 
or muon, high missing transverse momentum ($\met$) and at least three jets.
Similar searches in this channel have been published by the CDF and D0 collaborations~\cite{cdf_tprime,d0_tprime};
the most stringent limits preclude the existence of a $t^\prime$ quark with a mass below 358~GeV at 95\% confidence level (C.L.).
A search for $t^\prime \bar{t^\prime}$ in the dilepton final state has been performed by the ATLAS collaboration~\cite{atlas_tprime_dil},
excluding a $t^\prime$ quark with a mass below 350~GeV at 95\% C.L. The lepton+jets signature has also been recently exploited 
by the ATLAS collaboration to search for $b^\prime \bar{b^\prime}\to W^- t W^+ \bar{t}$~\cite{atlas_bprime_ljets}.

The ATLAS detector~\cite{atlas_det} consists of an inner tracking system surrounded by a 
superconducting solenoid, electromagnetic and hadronic calorimeters, and a muon spectrometer (MS).
The inner detector is immersed in a 2 T axial magnetic field, and consists of pixel and silicon microstrip detectors inside a transition radiation tracker, 
providing charged particle tracking in the region $|\eta|<2.5$~\footnote{Pseudorapidity is defined as $\eta = -\ln[\tan(\theta/2)]$, where
$\theta$ is the polar angle relative to the beam direction and
$\phi$ is the azimuthal angle in the plane transverse to the
beam direction.}.
The electromagnetic (EM) calorimeter is based on lead/liquid-argon (LAr).
Hadron calorimetry is based on two different detector technologies, with
scintillator tiles or LAr as active media, and with either steel, copper, or
tungsten as the absorber material. The calorimeters provide coverage
up to $|\eta|<4.9$.
The MS consists of superconducting air-core toroids, a system of trigger chambers covering
the range $|\eta|<2.4$, and high-precision tracking chambers allowing muon momentum measurements within
$|\eta|<2.7$.

The data set used in this analysis was recorded between March and June 2011 using single electron and muon triggers and 
includes only events collected under stable beam conditions and for which all detector 
subsystems were fully operational. The corresponding integrated luminosity is 1.04 fb$^{-1}$.
The event selection criteria 
closely follow those used in recent ATLAS top quark studies, e.g. Ref.~\cite{ttbar_atlas}.
Electron candidates are required to satisfy $\pt >25$~GeV and $|\eta|<2.47$, excluding the transition
region $1.37<|\eta|<1.52$ between the barrel and endcap EM calorimeters. 
Muon candidates are required
to satisfy $\pt >20$~GeV and $|\eta|<2.5$. 
The $\pt$ threshold requirement ensures that the selected leptons are in the efficiency plateau of the single-lepton triggers. 
Background from multi-jet production 
is suppressed by a requirement of $\met >35(20)$~GeV~\cite{ttbar_3pb} in the electron (muon) channel, followed by 
$\met +m_{\rm T}>60$~GeV, where $m_{\rm T}$ is the transverse mass of the lepton and 
$\met$~\footnote{The transverse mass is defined by the formula
$m_{\rm T} = \sqrt{2 p^\ell_{\rm T} \met (1-\cos\Delta\phi)}$, where
$p^\ell_{\rm T}$  is the $\pt$ of the lepton and $\Delta\phi$ is the
azimuthal angle separation between the lepton and $\met$.}.
The $\met$ is constructed from the vector sum of all calorimeter cells contained in 
topological clusters~\cite{topocluster}, calibrated at the energy scale of the associated high-$\pt$ object, 
and including contributions from selected muons.
Further requirements are that there be at least three jets with $\pt >25$~GeV and $|\eta|<2.5$, with at least one jet satisfying $\pt >60$~GeV.
Jets are reconstructed with the anti-$k_t$ algorithm~\cite{antikt} with radius parameter $R=0.4$, from
topological clusters of energy deposits in the calorimeters calibrated at the EM scale.
These jets are then calibrated to the particle level~\cite{jes} using a $\pt$- and $\eta$-dependent correction
factor derived from simulated events and validated using data.  
Finally, to further reduce the backgrounds, at least one jet is required to be identified as originating from 
the hadronization of a $b$ quark ($b$-tagging). 
This is achieved via an algorithm~\cite{btagging} using multivariate techniques to combine information from the impact parameters of displaced tracks as well 
as topological properties of secondary and tertiary decay vertices reconstructed within the jet;
a working point is used with $\sim 70\%$ efficiency for $b$-quark jets and a rejection factor of $\sim 100$ for jets originating from light quarks ($u$, $d$, $s$) or gluons.
Events with exactly one electron or one muon, and with exactly three jets or with four or more jets are analyzed separately to take 
advantage of their different  signal-to-background ratio and background composition, as discussed below.

\begin{table*}
\begin{center}
\begin{tabular}{lD{,}{\,\pm\,}{-1}D{,}{\,\pm\,}{-1}D{,}{\,\pm\,}{-1}D{,}{\,\pm\,}{-1}}
\hline
\hline
 &  \multicolumn{1}{c}{$e$+3 jets} & \multicolumn{1}{c}{$\mu$+3 jets} &  \multicolumn{1}{c}{$e$+$\geq$4 jets} & \multicolumn{1}{c}{$\mu$+$\geq$4 jets}\\
\hline
$t\bar{t}$& 2320 , 460 & 3000 , 630 & 4470 , 920 & 5900 , 1200 \\
$W$+jets& 1440 , 790 & 2200 , 1200 & 830 , 580 & 1160 , 790 \\
$Z$+jets& 92 , 53 & 118 , 62 & 86 , 56 & 83 , 46 \\
Single top& 382 , 68 & 554 , 94 & 262 , 70 & 325 , 79 \\
Dibosons& 28 , 7 & 37 , 11 & 12 , 5 & 17 , 5 \\
Multi-jet & 520 , 520 & 550 , 550 & 320 , 320 & 340 , 340 \\
\hline
Total prediction & 4800 , 1000 & 6500 , 1500 & 6000 , 1100 & 7800 , 1400 \\
Data& \multicolumn{1}{c}{4533} & \multicolumn{1}{c}{6421} & \multicolumn{1}{c}{6145} & \multicolumn{1}{c}{8149} \\
\hline
$t^\prime\bar{t^\prime}(400~\mathrm{GeV})$& 20.0 , 3.3 & 21.0 , 3.6 & 102.0 , 10.5 & 98.1 , 11.1 \\
\hline
\hline
\end{tabular}
\caption{\small{Number of events observed compared to the background expectation after
final event selection in each of the four channels considered. Also shown
are the expected signal yields assuming $m_{t^\prime}=400$~GeV. The quoted uncertainties are prior to the fit to data and
include both statistical and systematic contributions, taking into account correlations among processes.}}
\label{tab:yields}
\end{center}
\end{table*}

After event selection the main background is $t\bar{t}$ production, followed
by the production of a $W$ boson in association with jets ($W$+jets).
Smaller contributions arise from multi-jet events, single top quark, $Z$+jets and diboson production.
All of the backgrounds which do not involve top quarks are significantly suppressed by the $b$-tagging requirement.
Multi-jet events contribute to the selected sample via the misidentification of a jet or a photon as an electron
or the presence of a non-prompt lepton, e.g.~from a semileptonic $b$- or $c$-hadron decay.
The normalization and shape of the multi-jet background kinematic distributions are estimated via data-driven 
methods~\cite{ttbar_3pb}. 
For the $W$+jets background, the shape is estimated from the simulation but the normalization is estimated from the asymmetry between
$W^+$+jets and $W^-$+jets production~\cite{topasym} in data. 
All other backgrounds, as well as the signal, are estimated from the simulation and normalized
to their theoretical cross sections.
A summary of the background estimates in each of the four channels analyzed, and a comparison with the observed yields in data are
presented in Table~\ref{tab:yields}, showing a good agreement within the uncertainties.

Monte Carlo (MC) samples of  $t\bar{t}$ and single top quark background 
are generated using {\sc MC@NLO} v3.41~\cite{mcatnlo_1,*mcatnlo_2,*mcatnlo_3}, assuming a top quark mass of 172.5~GeV, using the {\sc CTEQ6.6} set of parton distribution functions (PDF)~\cite{cteq6}, and 
are normalized to the approximate next-to-next-to-leading-order 
(NNLO) theoretical cross sections~\cite{ttbarxs,stopxs,*stopxs_2}. 
Samples of $W/Z$+jets background are generated using {\sc Alpgen v2.13}~\cite{alpgen} and the {\sc CTEQ6L1} PDF set~\cite{cteq6}. 
The $Z$+jets background is normalized to the NNLO theoretical cross section~\cite{vjetsxs}, while the $W$+jets background normalization 
is extracted from data.
Both {\sc MC@NLO} and  {\sc Alpgen} are interfaced to  {\sc Herwig} v6.5~\cite{herwig} to model the parton shower and fragmentation, 
while {\sc Jimmy}~\cite{jimmy} is used to simulate the underlying event. The diboson backgrounds are modeled using {\sc Herwig} v6.5
and normalized to their NLO theoretical cross sections~\cite{mcfm}.  The signal is modeled using {\sc Pythia} 6.421~\cite{py6}.
Signal samples are generated for a range of masses, $m_{t^\prime}$, from 250 to 500~GeV in steps of 50~GeV and
are normalized to the approximate NNLO theoretical cross sections~\cite{ttbarxs} using the {\sc CTEQ6.6} PDF.
The MC samples generated using {\sc Herwig} or {\sc Pythia} use the {\sc MRST2007 LO*} PDF set~\cite{mrst}. 
All MC samples include multiple $pp$ interactions and are processed through a full simulation~\cite{atlas_sim}
of the detector geometry and response using {\sc Geant4}~\cite{geant}, and the same reconstruction software
as the data. Simulated events are corrected to match the object identification efficiencies and resolutions
determined in data control samples. The total signal detection efficiency, considering both lepton flavors and jet multiplicities
analyzed, ranges from 5.2\% for $m_{t^\prime}=250$~GeV to 17.3\% for $m_{t^\prime}=500$~GeV.

This analysis uses the reconstructed heavy quark mass ($m_{\rm reco}$) as the primary discriminating variable. 
In the case of events with $\geq 4$ jets, $m_{\rm reco}$ is estimated by performing
a kinematic likelihood fit~\cite{topasym} to the $t^\prime \bar{t^\prime} \to W^+b W^- \bar{b} \to \ell \nu b q\bar{q}^\prime \bar{b}$ hypothesis,
imposing the constraints that  $t^\prime$ and $\bar{t^\prime}$ have the same mass, and that the mass of the lepton-neutrino
system, as well as that of a jet pair, equals the nominal $W$ boson mass. The final state objects considered are the lepton, 
$\met$ and the four jets with highest $\pt$. Among all possible jet-parton permutations, the
one yielding the highest likelihood value after maximization over the fit parameters is kept. In the case of events 
with exactly three jets, $m_{\rm reco}$ is taken to
be the invariant mass of the three-jet system. In order to ensure a robust background prediction in the tail of the $m_{\rm reco}$ distribution,
a dynamic binning scheme is adopted;
starting from the high side and low side of the distributions, bins are merged until the statistical uncertainty
in the sum of the background predictions in that bin drops below $5\%$. 

Systematic uncertainties affecting the normalization and shape of the $m_{\rm reco}$ distribution are estimated
for both signal and background, taking into account correlations among processes as well as channels.
The dominant sources of uncertainty arise from the modeling of the $t\bar{t}$ background.  
The uncertainties on the $t\bar{t}$ background come from the theoretical 
uncertainty on the  cross section ($^{+7.0}_{-9.6}$\%) as well as the effects on both normalization and shape of the
$m_{\rm reco}$ distribution from a number of sources;
these are uncertainties 
on the fragmentation model (based on the comparison of {\sc Herwig} and {\sc Pythia} fragmentations), 
on the NLO event generator (based on the comparison of  {\sc MC@NLO} and {\sc Powheg}~\cite{powheg}) 
and on the top quark mass (taken to be $\pm 1$~GeV).

The uncertainty on the jet energy scale affects the normalization of signal (2--12\%) and backgrounds (5--30\%) modeled through the simulation,
as well as the shape of their  $m_{\rm reco}$ distributions.

Uncertainties on the modeling of initial- and final-state QCD radiation (ISR/FSR), evaluated by varying corresponding generator parameters, are considered as correlated between the $t\bar{t}$ 
background and the $t^\prime \bar{t^\prime}$ signal.

While the normalization is obtained from the asymmetry measurement, the uncertainties on the normalization of the $W$+jets background are derived from measurements of $W$+2 jets dominated data samples
and take into account the uncertainty on the heavy-flavor content of the samples as well as the extrapolation to 
higher jet multiplicities. The total uncertainty on the $W$+jets normalization is 50\% and 70\% for events with
exactly 3 jets and $\geq$4 jets, respectively. Uncertainties on the shape of the $m_{\rm reco}$ distribution for
the $W$+jets background are estimated by varying the choices of the matching scale (from 15 to 10~GeV) and 
the factorization scale (from $\mu_\mathrm{F}^2=m_W^2 + \sum p_{\rm T, jet}^2$ to $\mu_\mathrm{F}^2=m_W^2 + p_{{\rm T},W}^2$) in {\sc Alpgen}.

Uncertainties on the modeling of the $b$-tagging algorithms affect the identification of $b/c$-jets 
(6--8\% for signal and backgrounds containing top quarks, 6--12\% for the other backgrounds) as well as the mis-identification 
of light jets ($< 0.5\%$ for signal and backgrounds containing top quarks and up to 5\% for the other backgrounds).
The $Z$+jets, single top and diboson backgrounds are varied within the uncertainty on their theoretical cross sections.
The uncertainty on the multi-jet background event normalizations is conservatively taken as 100\%.
Uncertainties on the shapes of the multi-jet background are derived by varying the lepton identification
criteria used to extract this background.

The uncertainties on the lepton identification and trigger efficiencies, as well as their energy scales and resolutions, impact the yields by 3\% for electrons and 6\% for muons.

Uncertainties on the integrated luminosity (3.7\%)~\cite{lumi,*lumi_old}, jet reconstruction efficiency, jet resolution modeling, effect of multiple $pp$ interactions on the modeling of 
the $\met$ and treatment of imperfections in the detector description in the MC simulation are also considered and are all found to have a very small effect on the result.

Good agreement between the data and the background prediction is observed both in terms of overall normalization and
shape of the $m_{\rm reco}$ distribution. The $m_{\rm reco}$ distribution
is analyzed using a log-likelihood ratio $LLR=-2\log (L_{\rm s+b}/L_{\rm b})$ as test-statistic, where $L_{\rm s+b}$ ($L_{\rm b}$)
is a Poisson likelihood to observe the data under the signal-plus-background (background-only) hypothesis.
The per-bin signal and background predictions are parameterized in terms of 12 nuisance parameters, describing
the effect of leading sources of systematic uncertainty such as jet energy scale, ISR/FSR, and $t\bar{t}$, $W$+jets and QCD multi-jet normalizations. 
The impact of systematic uncertainties on the sensitivity of the search is reduced by maximizing both
likelihood functions, $L_{\rm s+b}$ and $L_{\rm b}$, with respect to these nuisance parameters,
subject to Gaussian constraints of their prior values. The set of fitted nuisance parameters is chosen based
on their overall impact on the search sensitivity, the expected constraining power of the data and their suitability to be treated
as continuous parameters. 
The simultaneous constraint of several of these systematic uncertainties is possible because of the inclusion of the 3-jet channel in the analysis.
The latter has a higher fraction of $W$+jets background than the $\geq$4-jets channel, and provides sensitivity to event migration to different jet multiplicities
when varying uncertainties such as jet energy scale or ISR/FSR.
In addition to the jet multiplicity spectrum, the jet energy scale affects the peak position of the $m_{\rm reco}$ spectrum for
$t\bar{t}$ background, and can be constrained owing to the small uncertainty on the measured top quark mass~\cite{PDG}.
Nuisance parameters associated with smaller systematic uncertainties (e.g. lepton identification/trigger) are 
only weakly constrained.

\begin{figure}
\centering
\subfigure[a][]{\includegraphics[width=0.45\textwidth]{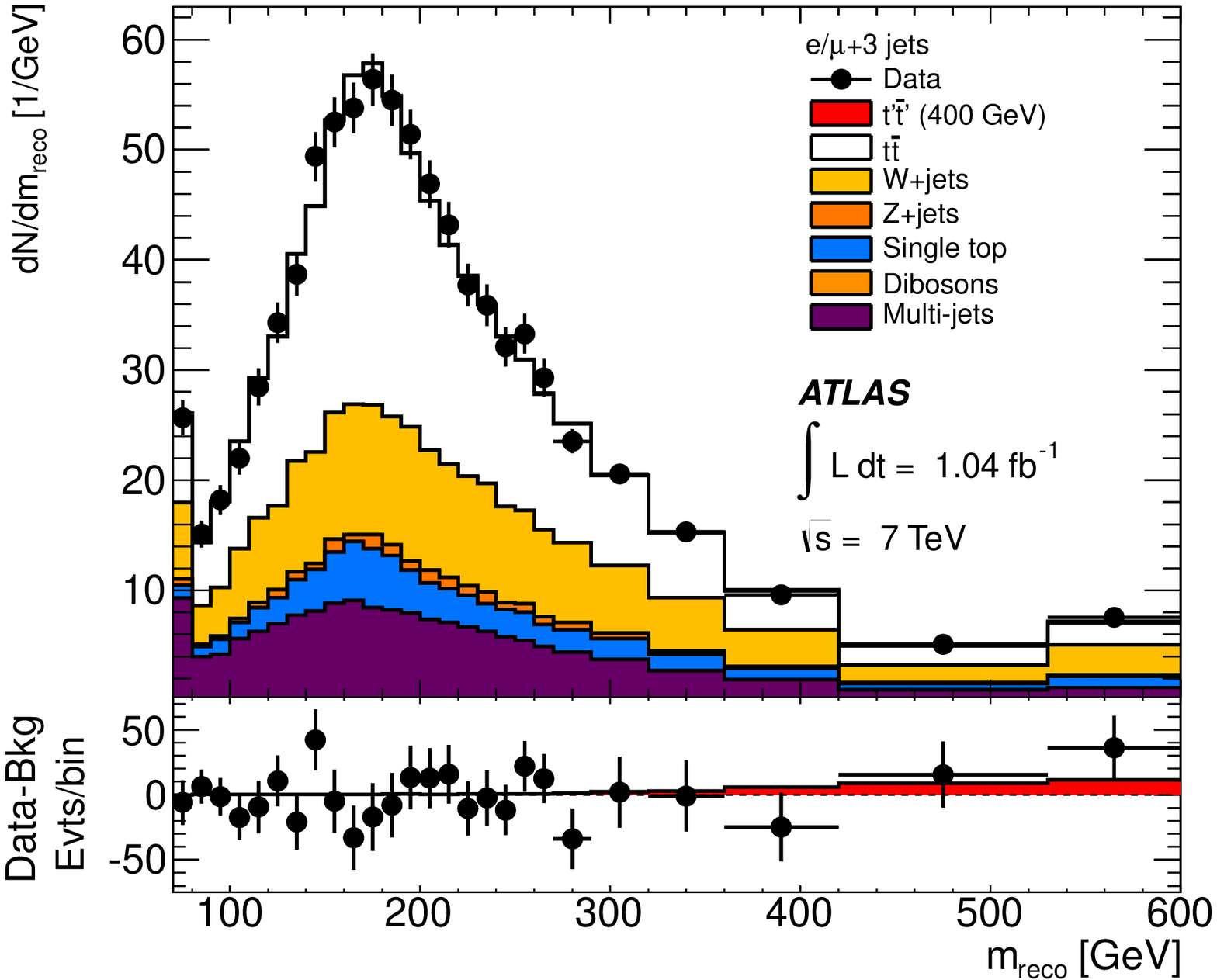}} 
\subfigure[b][]{\includegraphics[width=0.45\textwidth]{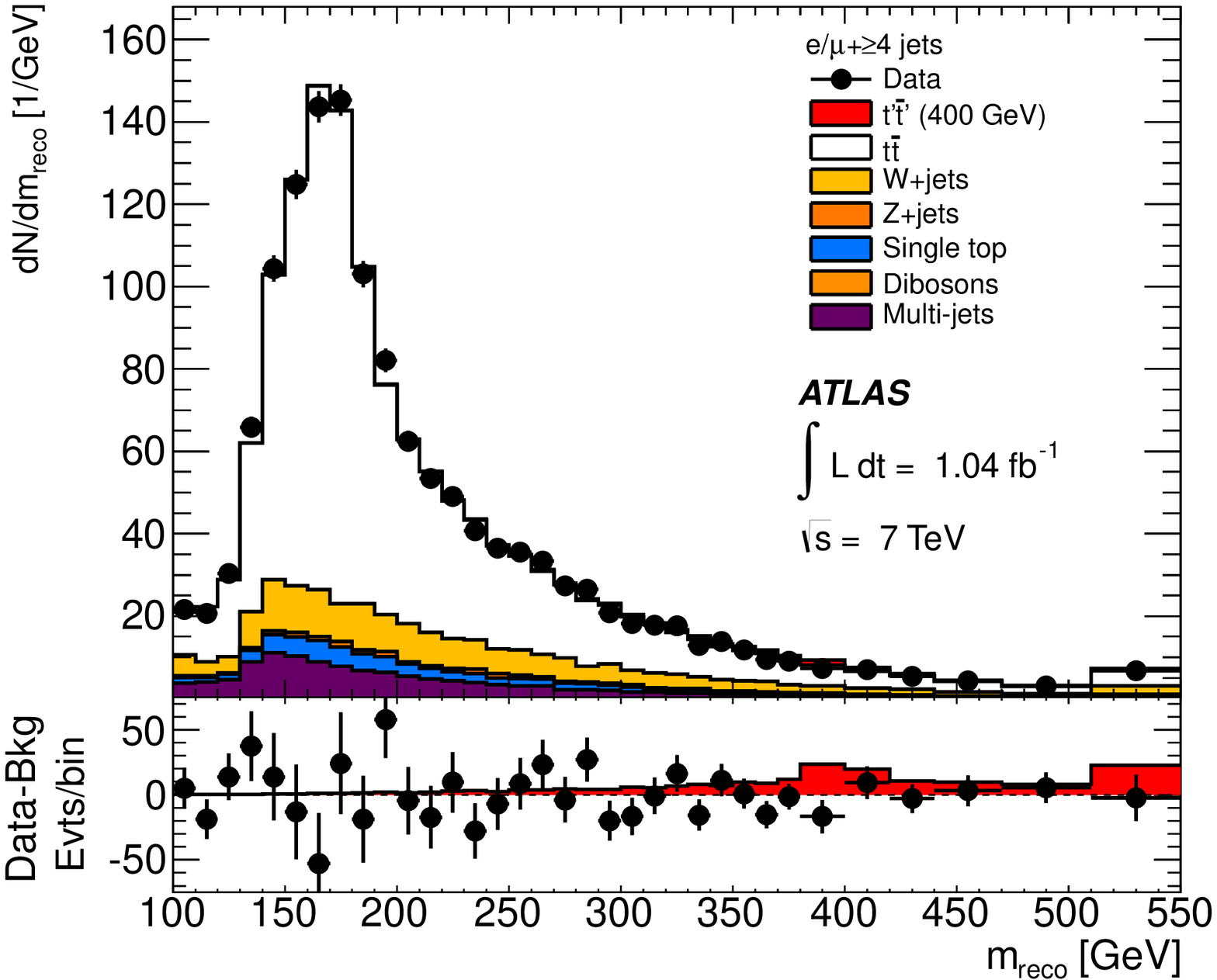}} 
\caption{$m_{\rm reco}$ distribution in the combined (a) $e/\mu$+3 jets and (b) $e/\mu$+$\geq$4 jets channels. The data (points) are compared to the SM background 
predictions using the values of the nuisance parameters obtained from the fit to data under the background-only hypothesis (stacked histograms). 
In the top panels the bin contents have been divided by bin width.
The bottom panels show the background-subtracted data distribution. 
The underflow and overflow have been folded into the first and last bins, respectively.
Also shown is the expected contribution from a signal with mass $m_{t^\prime}=400$~GeV (histogram). }
\label{fig:mjjj_postfit}
\end{figure}

Figure~\ref{fig:mjjj_postfit} shows a comparison
of the post-fit $m_{\rm reco}$ distribution between data and the background prediction for the combined $e/\mu+3$ jets and $e/\mu + \geq 4$ jets channels. 
The fitted parameters are typically within one standard deviation of their nominal values and their uncertainties are consistent with 
expectations based on pseudo-experiments. 
Several additional studies were performed to check the integrity of the 
fitting procedure. The likelihood was verified to be parabolic near the minimum for each of the fitted parameters and 
to yield reasonable fit uncertainties;  the lack of sensitivity to the assumed $\pt$ and $\eta$ correlation of the jet energy scale
uncertainty was verified.

In the absence of any significant data excess, either in the $e$+jets or $\mu$+jets channels
individually or in their combination, 95\% C.L. upper limits on the $t^{\prime} \bar{t^\prime}$ production cross
section are derived using the $CL_{\rm S}$ method~\cite{cls,*cls_2}, which employs the $LLR$ test-statistic
described above. Pseudo-experiments are generated under both the signal-plus-background
(s+b) and background-only (b) hypotheses, taking into account per-bin statistical fluctuations
of the total predictions according to Poisson statistics, as well as Gaussian fluctuations
in the signal and background expectations describing the effect of systematic uncertainties.
The fraction of s+b and b pseudo-experiments with $LLR$ larger than the median or observed $LLR$
defines $CL_{\rm s+b}$ and $CL_{\rm b}$ for the expected or observed limits,
respectively.  Signal cross sections for which $CL_{\rm s}=CL_{\rm s+b}/CL_{\rm b}<0.05$ are deemed excluded at
the 95\% C.L.

\begin{figure}
\centering
\includegraphics[width=0.48\textwidth]{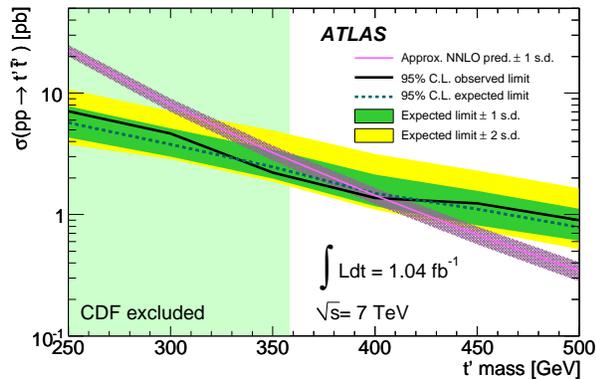}
\caption{Observed (solid line) and expected (dashed line) 95\% C.L. upper limits on the $t^\prime \bar{t^\prime}$ cross section 
as a function of the $t^{\prime}$ mass. The surrounding shaded bands correspond to the 1 and 2 standard deviations (s.d.) around the expected limit. 
The thin line shows the theoretical prediction including its 1 s.d. uncertainty band. The shaded area is the 
mass region previously excluded by the CDF experiment~\cite{cdf_tprime}.}
\label{fig:limits}
\end{figure}

The resulting observed and expected upper limits on the $t^{\prime} \bar{t^\prime}$ production cross section are shown in Fig.~\ref{fig:limits} as a function 
of the $t^\prime$ mass, compared to the theoretical prediction, assuming a $BR(t^\prime \rightarrow Wb)=1$.
As a result,  an observed (expected) 95\%  C.L.  lower limit of 404 (394) GeV on the mass of the $t^\prime$ quark is derived.

In summary, a search for $t^\prime \bar{t^\prime}$ production has been performed in the lepton+jets final state under the assumption
$BR(t^\prime \to Wb)=1$.  No significant excess of events in the tail of the $m_{\rm reco}$ distribution was found, 
resulting in an observed lower  limit of  $m_{t^\prime}>404$~GeV at 95\% C.L. 
This represents the most stringent limit to date.
This limit is also directly applicable to a down-type vector-like quark with electric
charge of $-4/3$ decaying into a $W$ boson and a $b$ quark~\cite{jaas}.

\begin{acknowledgments}
We thank CERN for the very successful operation of the LHC, as well as the
support staff from our institutions without whom ATLAS could not be
operated efficiently.

We acknowledge the support of ANPCyT, Argentina; YerPhI, Armenia; ARC,
Australia; BMWF, Austria; ANAS, Azerbaijan; SSTC, Belarus; CNPq and FAPESP,
Brazil; NSERC, NRC and CFI, Canada; CERN; CONICYT, Chile; CAS, MOST and NSFC,
China; COLCIENCIAS, Colombia; MSMT CR, MPO CR and VSC CR, Czech Republic;
DNRF, DNSRC and Lundbeck Foundation, Denmark; ARTEMIS and ERC, European Union;
IN2P3-CNRS, CEA-DSM/IRFU, France; GNAS, Georgia; BMBF, DFG, HGF, MPG and AvH
Foundation, Germany; GSRT, Greece; ISF, MINERVA, GIF, DIP and Benoziyo Center,
Israel; INFN, Italy; MEXT and JSPS, Japan; CNRST, Morocco; FOM and NWO,
Netherlands; RCN, Norway; MNiSW, Poland; GRICES and FCT, Portugal; MERYS
(MECTS), Romania; MES of Russia and ROSATOM, Russian Federation; JINR; MSTD,
Serbia; MSSR, Slovakia; ARRS and MVZT, Slovenia; DST/NRF, South Africa;
MICINN, Spain; SRC and Wallenberg Foundation, Sweden; SER, SNSF and Cantons of
Bern and Geneva, Switzerland; NSC, Taiwan; TAEK, Turkey; STFC, the Royal
Society and Leverhulme Trust, United Kingdom; DOE and NSF, United States of
America.

The crucial computing support from all WLCG partners is acknowledged
gratefully, in particular from CERN and the ATLAS Tier-1 facilities at
TRIUMF (Canada), NDGF (Denmark, Norway, Sweden), CC-IN2P3 (France),
KIT/GridKA (Germany), INFN-CNAF (Italy), NL-T1 (Netherlands), PIC (Spain),
ASGC (Taiwan), RAL (UK) and BNL (USA) and in the Tier-2 facilities
worldwide.

\end{acknowledgments}

\bibliography{EXOT-2011-03}
\onecolumngrid

\clearpage
\begin{flushleft}
{\Large The ATLAS Collaboration}

\bigskip

G.~Aad$^{\rm 48}$,
B.~Abbott$^{\rm 110}$,
J.~Abdallah$^{\rm 11}$,
A.A.~Abdelalim$^{\rm 49}$,
A.~Abdesselam$^{\rm 117}$,
O.~Abdinov$^{\rm 10}$,
B.~Abi$^{\rm 111}$,
M.~Abolins$^{\rm 87}$,
O.S.~AbouZeid$^{\rm 157}$,
H.~Abramowicz$^{\rm 152}$
% ...
\bigskip

% ...
$^{10}$ Institute of Physics, Azerbaijan Academy of Sciences, Baku, Azerbaijan\\
$^{11}$ Institut de F\'isica d'Altes Energies and Departament de F\'isica de la Universitat Aut\`onoma  de Barcelona and ICREA, Barcelona, Spain\\
% ...
$^{48}$ Fakult\"{a}t f\"{u}r Mathematik und Physik, Albert-Ludwigs-Universit\"{a}t, Freiburg i.Br., Germany\\
$^{49}$ Section de Physique, Universit\'e de Gen\`eve, Geneva, Switzerland\\
% ...
$^{87}$ Department of Physics and Astronomy, Michigan State University, East Lansing MI, United States of America\\
% ...
$^{110}$ Homer L. Dodge Department of Physics and Astronomy, University of Oklahoma, Norman OK, United States of America\\
$^{111}$ Department of Physics, Oklahoma State University, Stillwater OK, United States of America\\
% ...
$^{117}$ Department of Physics, Oxford University, Oxford, United Kingdom\\
% ...
$^{152}$ Raymond and Beverly Sackler School of Physics and Astronomy, Tel Aviv University, Tel Aviv, Israel\\
% ...
$^{157}$ Department of Physics, University of Toronto, Toronto ON, Canada\\
% ...
\end{flushleft}


\end{document}